\documentclass[aps,prb,longbibliography,twocolumn,superscriptaddress]{revtex4-2}
\usepackage{amsmath,amssymb,amsfonts,bm}
\usepackage{graphicx}
\usepackage{epstopdf}
\usepackage{dcolumn}
\usepackage{mathrsfs}
\usepackage{bbold}
\usepackage{dsfont}
\usepackage{float}
\usepackage{ulem}
\usepackage[colorlinks=true,linkcolor=blue,citecolor=blue, urlcolor=blue,bookmarks=false]{hyperref}

\usepackage{tgtermes}
\usepackage{multirow}

\begin{document}
\title{ Majorana corner modes and flat-band Majorana edge modes in superconductor/topological-insulator/superconductor junctions}
\author{Xiao-Ting Chen}
\thanks{These authors contributed equally to this work.}
\affiliation{Institute for Advanced Study and School of Physical Science and Technology, Soochow University, Suzhou 215006, China.}
\author{Chun-Hui Liu}
\thanks{These authors contributed equally to this work.}
\affiliation{Beijing National Laboratory for Condensed Matter Physics, Institute of Physics, Chinese Academy of Sciences, Beijing 100190, China}
\affiliation{School of Physical Sciences, University of Chinese Academy of Sciences, Beijing 100049, China}
\author{Dong-Hui Xu}\email[]{donghuixu@cqu.edu.cn}
\affiliation{Department of Physics and Chongqing Key Laboratory for Strongly Coupled Physics, Chongqing University, Chongqing 400044, China}
\affiliation{Center of Quantum Materials and Devices, Chongqing University, Chongqing 400044, China}
\author{Chui-Zhen Chen}\email[]{czchen@suda.edu.cn}
\affiliation{Institute for Advanced Study and School of Physical Science and Technology, Soochow University, Suzhou 215006, China.}
\begin{abstract}
  Recently, superconductors with higher-order topology have stimulated extensive attention and research interest. Higher-order topological superconductors exhibit unconventional bulk-boundary correspondence, thus allow exotic lower-dimensional boundary modes, such as Majorana corner and hinge modes. However, higher-order topological superconductivity has yet to be found in naturally occurring materials. In this work, we investigate higher-order topology in a two-dimensional Josephson junction comprised of two $s$-wave superconductors separated by a topological insulator thin film. We found that zero-energy Majorana corner modes, a boundary fingerprint of higher-order topological superconductivity, can be achieved by applying magnetic field. When an in-plane Zeeman field is applied to the system, two corner modes appear in the superconducting junction. Furthermore, we also discover a two dimensional nodal superconducting phase which supports flat-band Majorana edge modes connecting the bulk nodes. Importantly, we demonstrate that zero-energy Majorana corner modes are stable when increasing the thickness of topological insulator thin film.

\end{abstract}

\maketitle

\section{Introduction}
The search for topological superconductors which host Majorana zero-energy
modes has been one of the central subjects in condensed matter physics \cite{kane2010rmp,qi2011topological,alicea2012new,beenakker2013search,lutchyn2018majorana,flensberg2021engineered},
since they provide an ideal platform to potential applications in quantum computations
based on non-Abelian statistics \cite{ivanov2001non,kitaev2003fault,nayak2008non,Alicea_2011,flensberg2011non,van_Heck_2012,Aasen_2016}.
In 2001, A.Y. Kitaev proposed to realize Majorana zero-energy modes at the ends of one-dimensional $p$-wave superconductors \cite{kitaev2001unpaired}.
Experimentally, the signatures of zero-energy modes were reported to been observed in spin-orbital coupled semiconductor wires in proximity to $s$-wave superconductors \cite{Mourik_2012,Das2012,Deng2012,lutchyn2010}
and in the vortex of the iron-based superconductor \cite{Zhang2018V,Kong2019,Gao2018,Feng2018,zhu2020nearly}. However, conclusive identification of Majorana zero energy modes and scalable fabrication of Majorana networks remain challenging \cite{lutchyn2018majorana,flensberg2021engineered,karzig2017scalable}.

Recently, higher-order topological phases of matter, such as the higher-order topological
insulators and higher-order topological superconductors have been identified as a novel topological
state, which feature the unconventional bulk-boundary correspondence \cite{peng2017boundary,song2017d,langbehn2017reflection,benalcazar2017electric,benalcazar2017quantized,schindler2018higher,van2018higher,ezawa2018strong,franca2018anomalous,ezawa2018magnetic,ezawa2018higher,ezawa2018topological,kunst2018lattice,you2018higher,wang2019higher,kudo2019higher,trifunovic2019higher,yue2019symmetry,xu2019higher,hua2020higher,chen2020higher,rasmussen2020classification}.
Generally, the $n$th higher-order topological superconductors in $d$ dimensions support $(d-n)$ dimensional gapless boundary excitations with $2\le n\le d$ \cite{langbehn2017reflection,geier2018second,yan2018Majorana,wang2018high,zhu2018Tunable,khalaf2018higher,wang2018weak,khalaf2018higher,liu2018Majorana,hsu2018majorana,shapourian2018topological,volpez2019second,Pan2019lattice,franca2019phase,Varjas2019PRL,yan2019higher,laubscher2020kramers,ZengPRL19,zhu2019second,Bultinck2019PRB,zhang2019helical,ZhangRX19PRL2,wu2019higher,ghorashi2019second,yan2019majorana,liu2021topological,hsu2020inversion,Wu2020boundary,Kheirkhah2020first,laubscher2020majorana,wu2020plane,ahn2020higher,roy2020higher,ghorashi2020vortex,tiwari2020chiral,fu2021chiral,kheirkhah2020majorana,Ikegaya2021tunable,roy2021mixed,qin2022topological,Scammell2022intrinsic},
which is in contrast to the $d$ dimensional conventional topological superconductors with $(d-1)$ dimensional boundary excitations \cite{kane2010rmp,qi2011topological,shen2012topological,bernevig2013topological}.
In this endeavor, Majorana zero-energy modes are supposed to be localized at the corners of a two dimensional
second-order topological superconductor (SOTSC)~\cite{langbehn2017reflection,yan2018Majorana,wang2018high,zhu2018Tunable,liu2018Majorana,Pan2019lattice,Wu2020boundary,wu2020plane}. Due to naturally occurring topological superconductors are extremely rare, SOTSCs with Majorana Kramers pairs of zero modes or single Majorana zero mode at a corner have been proposed in artificial materials, such as a quantum spin Hall insulators (QSHI) in proximity to a $d$-wave or an $s_\pm$-wave superconductor~\cite{yan2018Majorana,wang2018high}, two coupled chiral $p$-wave superconductors~\cite{zhu2018Tunable}, Rashba spin-orbit coupled $\pi$-junction \cite{volpez2019second}, etc. However, the experimental implementation remains challenging because
of the requirements of ideal helical-edge modes of the QSHI, unconventional superconductivities or complicated junction. 
Fortunately, topological insulator thin films/$s$-wave superconductors hybrid structures have been successfully 
fabricated~\cite{wang2012coexistence,xu2014momentum,xu2014artificial} and used to engineer the first-order topological 
superconductors~\cite{Fu2008superconducting,liu2011helical}. Even more importantly, a spin-selective Andreev reflection in the 
vortex of topological insulator/superconductor heterostructure was reported in a scanning tunnel microscope measurement~\cite{sun2016majorana}, which is regarded as a fingerprint of Majorana zero-energy modes. 

\begin{figure}[htbp]
	\centering
	\includegraphics[width=3.2in]{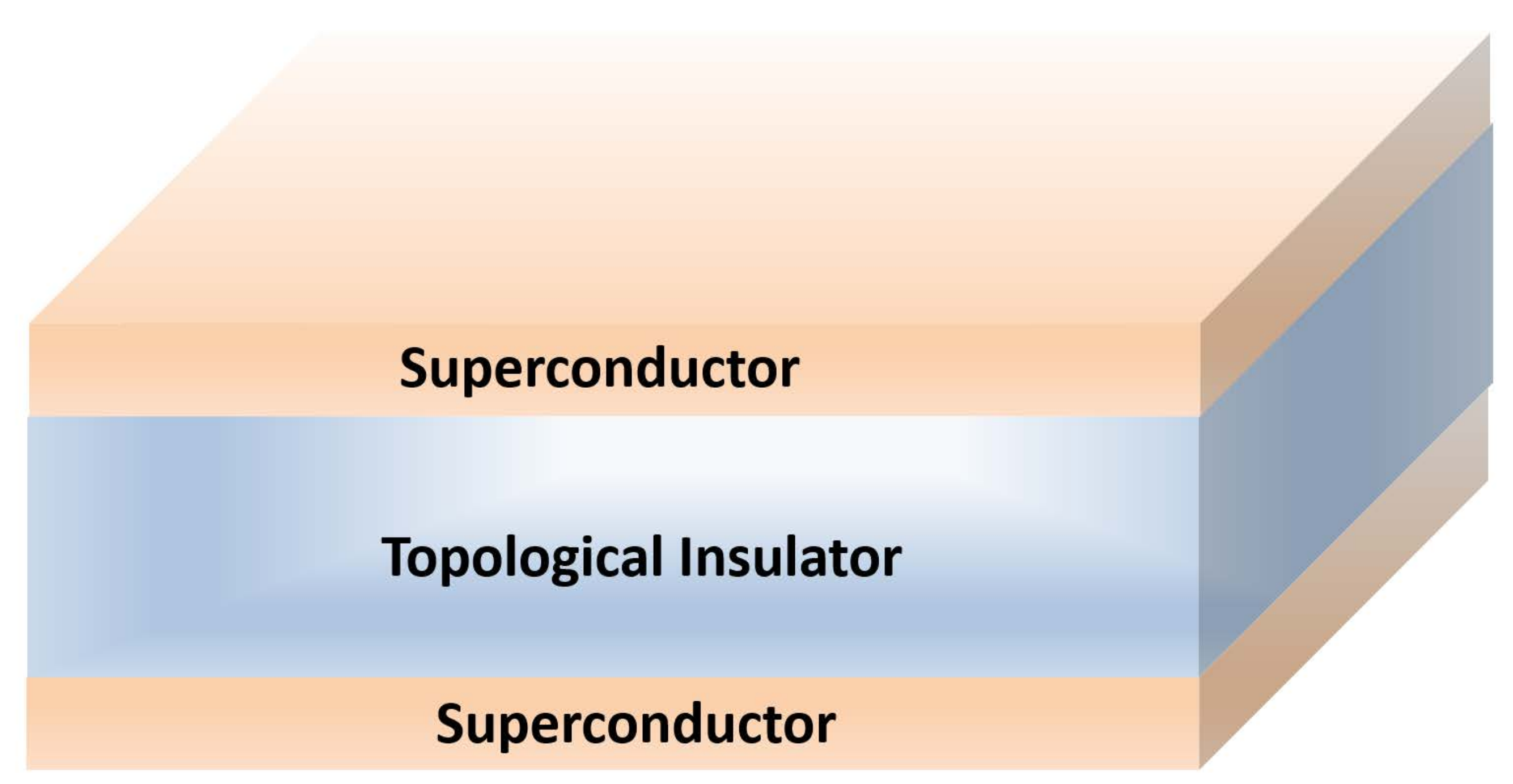}
	\caption{(Color online). Schematic plots of topological insulator multilayer based superconductor junction.
		\label{fig1} }
\end{figure}
In this work, we propose that a topological insulator thin film sandwiched between two $s$-wave superconductors with a phase difference $\pi$ (see Fig.~\ref{fig1})
can realize an SOTSC with two localized Majorana corner modes when applying an in-plane magnetic field. Moreover, a nodal topological superconducting phase hosting flat-band Majorana edge modes emerges when tuning the magnetic field. We also show that Majorana corner modes exist when varying the number of layers of topological insulator thin film. Our findings make the superconductor-topological-insulator-superconductor junctions an incredibly fertile platform for exploring topological superconducting phase.

The structure of this paper is as follows. In Sec.~\ref{secii} we describe
our setup, which consists of a topological insulator thin film, in proximity to a top and a
bottom superconductors with a phase difference of $\pi$, see
Fig.~\ref{fig1}. In Sec.~\ref{seciii}, we present the Majorana corner modes and flat-band Majorana edge modes, and plot a topological superconducting phase diagram. Meanwhile, we do the symmetry analysis and calculate topological invariants to characterize the topological superconducting phases. We also construct an edge theory based on the perturbation theory to elucidate the formation of Majorana corner modes. In Sec.~\ref{seciv}, We study how the thickness of topological insulator thin film affects the zero-energy Majorana corner modes. We summarize our results in Sec.~\ref{secv}.

\begin{figure*}[htbp]
	\centering
	\includegraphics[width=6.0in]{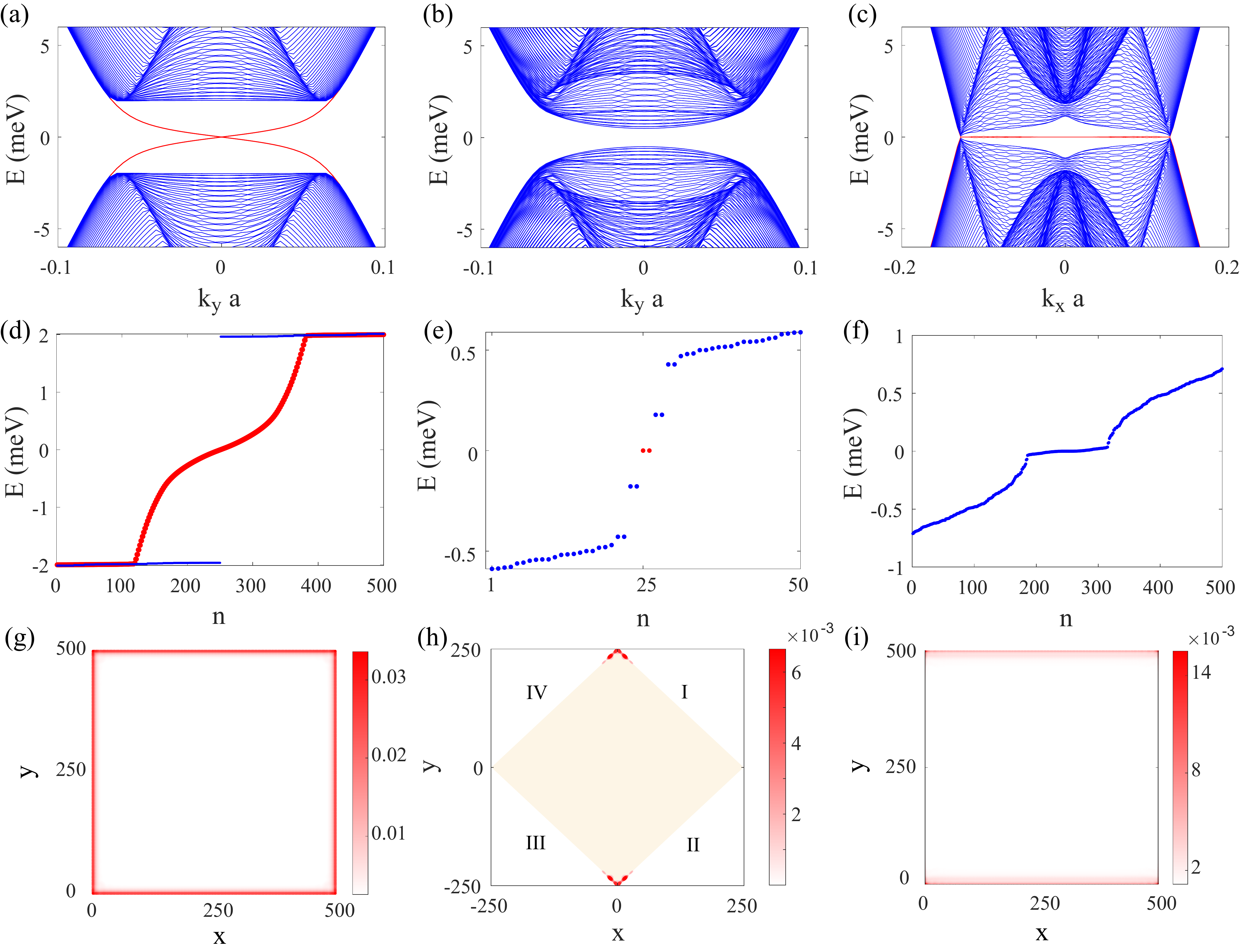}
	\caption
	{(Color online).
	Energy dispersion of a ribbon geometry for (a) $V_{x}=0\,\mathrm{meV}$,
(b) $V_{x}=1.5\,\mathrm{meV}$,
(c) $V_{x}=12\,\mathrm{meV}$.
In (a) and (b), we take the open boundary condition in the $x$-direction.
In (c), we take the open boundary condition in the $y$-direction as the bulk nodes are located along the $k_x$-axis.
(d) Bule (red) dots are energy spectrum of the configuration in (g) under the periodic boundary condition (open boundary condition) in all boundaries,
the parameters the same as (a).
(e) The energy spectrum of the configuration in (h) with the open boundary condition in all boundaries, the parameters the same as (b).
(f) The energy spectrum of configuration in (i) with the open boundary condition in the
$y$-direction and the periodic boundary condition in the $x$-direction, the parameters the same as (c).
 (g) Local density of states of gapless Majorana edge modes in (d).
 (h) Local density states of the ingap Majoarana bound states at zero energy marked in red in (e).
 (i) Local density of flat-band edge modes in (f).
 In (a)-(i), the parameters are $\mu=12\,\mathrm{meV}$, $\Delta_{}=2\,\mathrm{meV}$.
		\label{fig2} }
\end{figure*}

\section{Model} 
\label{secii}

In reciprocal space, the effective Hamiltonian of the superconductor/topological-insulator/superconductor heterostructure can be written as
\begin{equation}
\begin{split}
\label{Htd}
    H({\bf k})=&\frac{A}{a}\sin( k_{x}a)\sigma_{x}\tau_{z}+\frac{A}{a}\sin (k_{y}a)\rho_{z}\sigma_{y}\tau_{z}\\+&M(\mathbf{k})\rho_{z}\tau_{x}
    +\Delta\rho_{y}\sigma_{y}\tau_{z}+V_{x}\rho_{z}\sigma_{x}-\mu\rho_{z},
\end{split}
\end{equation}
in the Nambu basis $C_{\mathbf{k}}=(c_{\mathbf{k}l,\uparrow},c_{\mathbf{k}l,\downarrow},c_{-\mathbf{k}l,\uparrow}^{\dagger},c_{-\mathbf{k}l,\downarrow}^{\dagger})^{T}$, where $\uparrow$ and
$\downarrow$ represent two electron spin directions, and $l=1,2$ is the layer index.
$\sigma_{i},\tau_{i}$ and $\rho_{i}$ $(i=x,y,z)$ are the Pauli matrices. In Eq. (\ref{Htd}), they acting on the spin, layer, and particle-hole spaces, respectively.
$\sigma_{0},\tau_{0}$ and $\rho_{0}$ are the $2\times2$ identity matrices.
 $M({\bf k})=m_{0}-\frac{{2}m_{1}}{a^2}[2-\cos (k_{x}a)-\cos (k_{y}a)]$ describes the mass induced by the hybridization of the top and bottom surfaces of the topological insulator thin film. $A$ is the characteristic parameter of the kinetic energy of the Dirac fermions, and $\mu$ denotes the chemical potential. 
 $\Delta$ is the $s$-wave pairing amplitude, and the pairing functions of the top and bottom surfaces of the thin film have opposite sign, which makes the setup a Josephson junction with a $\pi$ phase shift. $V_{x}$ represents the in-plane Zeeman field applied along the $x$ direction.
We set the lattice constant $a=5\,\mathrm{nm}$, ${A}=300\, \mathrm{meV\cdot nm}$ and ${m_{1}}=150\, \mathrm{meV\cdot nm^{2}}$ \cite{Zhang_2010NP}.
For our purpose, we set $m_{0}=-2$ meV and $\Delta=2$ meV in this case.

Before turning to discuss the Majorana corner modes in this junction, we would like to give a brief discussion about the case without applied magnetic field. In this case, time-reversal symmetry restores, a topological superconducting phase with gapless helical Majorana edge modes [see Fig.~\ref{fig2}(a)] exists when $m_0^2<\Delta^2+\mu^2$ \cite{liu2011helical}. The helical edge modes are confirmed by calculating the Bogoliubov quasiparticle energy spectrum and the local density of states for a square nanodisk, which are shown in Figs.~\ref{fig2}(d) and \ref{fig2}(g).

\section{Majorana corner modes and flat-band edge modes} 
\label{seciii}
When turning on the in-plane Zeeman field, we find that the gapless helical Majorana edge modes are not stable owing to time-reversal symmetry breaking, in the meantime, an energy gap opens as shown in Fig. \ref{fig2}(b). This gap signals the occurrence of the SOTSC. We compute the Bogoliubov quasiparticle spectrum for a finite-sized sample with a rhombus geometry. The spectrum depicted in Fig.~\ref{fig2}(e) shows two degenerate Majorana ingap bound states at zero energy, which reside at the top and bottom corners of the rhombus, respectively, as depicted in Fig.~\ref{fig2}(h). These two Majorana corner modes are a smoking-gun signature of the SOTSC.  
In the following, we will construct a topological invariant and an edge theory based on the Jackiw-Rebbi mechanism to characterize the zero-energy Majorana corner modes in SOTSC phase. 

The Hamiltonian maintains the mirror symmetry: $C_{2y}H(k_x,k_y)C_{2y}^{-1}=H(-k_x,k_y)$ with $C_{2y}=i\sigma_{x}\tau_{x}$. Along the mirror invariant axis $k_{x}=0$ of the first Brillouin zone (BZ), $H(k_x=0,k_y)$ commutes with $C_{2y}$ operator. We can use a mirror winding number
along this axis to characterize the topological properties of the Majorana corner modes~\cite{park2019higher,liu2019second,liu2021topological}. The expression of $H(k_x=0,k_y)$ is
\begin{equation}
	\begin{split}
		\label{Hm}
		H(0,k_{y})=&\frac{A}{a}\sin(k_{y}a)\rho_{z}\sigma_{y}\tau_{z}+M(0,k_y)\rho_{z}\tau_{x}\\+&\Delta_{}\rho_{y}\sigma_{y}\tau_{z}+V_{x}\rho_{z}\sigma_{x}-
		\mu\rho_{z},
	\end{split}
\end{equation}
where $M(0,k_y)=m_{0}-\frac{2m_{1}}{a^{2}}[1-\cos(k_{y}a)]$.
$C_{2y}$ has two fourfold degenerate eigenvalues of $\pm1$.
The eigenvectors with eigenvalue of $+1$ are:
\begin{equation}
	\begin{split}
		\chi_{1}=&|\rho_{z}=1,\sigma_{x}=1,\tau_{x}=1\rangle, \\
		\chi_{2}=&|\rho_{z}=-1,\sigma_{x}=1,\tau_{x}=1\rangle,\\
		\chi_{3}=&|\rho_{z}=1,\sigma_{x}=-1,\tau_{x}=-1\rangle,\\
		\chi_{4}=&|\rho_{z}=-1,\sigma_{x}=-1,\tau_{x}=-1\rangle, \label{chic1}
	\end{split}
\end{equation}
which constitues the $+1$ eigenspace.
The eigenvectors with eigenvalue of $-1$ are:
\begin{equation}
	\begin{split}
		\chi_{5}=&|\rho_{z}=1,\sigma_{x}=1,\tau_{x}=-1\rangle, \\
		\chi_{6}=&|\rho_{z}=-1,\sigma_{x}=1,\tau_{x}=-1\rangle,\\
		\chi_{7}=&|\rho_{z}=1,\sigma_{x}=-1,\tau_{x}=1\text{\textrangle},\\
		\chi_{8}=&|\rho_{z}=-1,\sigma_{x}=-1,\tau_{x}=1\text{\textrangle}, \label{chic2}
	\end{split}
\end{equation}
which constitues the $-1$ eigenspace.
Projecting $H(0,k_{y})$ into
$+1$ eigenspace ($-1$ eigenspace) of $C_{2y}$, we can get a Hamiltonian in the subspace
$H_{+}(0,k_{y})$ [$H_{-}(0,k_{y})$]. Using $\chi_{1}, \chi_{2}, ..., \chi_{8}$
as a new basis set of Hilbert space, we can get
$H(0,k_{y})=H_{+}(0,k_{y})\text{\ensuremath{\bigoplus}}H_{-}(0,k_{y})$ with
\begin{equation}
	\begin{split}
		\label{Hm2}
		H_{\pm}(0,k_{y})=&\mp\frac{A}{a}\sin(k_{y}a)\rho_{z}\sigma_{y}+M(0,k_y)\rho_{z}\sigma_{z}\\&\mp\Delta_{}\rho_{y}\sigma_{y}\pm V_{x}\rho_{z}\sigma_{z}-\mu
		\rho_{z}.
	\end{split}
\end{equation}
In the first BZ, we can define
the Wilson loop operator $W_{\pm,k_{y}}$ of $H_{\pm}(0,k_{y})$, along the mirror-invariant axis $k_{x}=0$. The mirror winding number
$\nu_{\pm}$ can be written as \cite{park2019higher,liu2019second}:
\begin{equation}
	\label{vpm}
	\nu_{\pm}=\frac{1}{i\pi}\log(\det[W_{\pm,k_{y}}])\,\text{mod} \;2.
\end{equation}
When corner modes occur, the mirror winding number $\nu_{+}=\nu_{-}=1$ \cite{liu2021topological}.

Next, we use degenerate the perturbation theory to deduce the SOTSC phase. 
For simplicity, we only consider $\mu=0$, and $V_{x}$ terms are small enough such that
they can be treated as perturbations.
The low energy expansion of the SOTSC Hamiltonian in Eq.~(\ref{Htd}) at ${\bf k}=(0,0)$ is
\begin{equation}
	\begin{split}
	\label{Hl}
	H_{l}=&A k_{x}\sigma_{x}\tau_{z}+Ak_{y}\rho_{z}\sigma_{y}\tau_{z}+[m_0-m_1(k_x^2+k_y^2)]\rho_{z}\tau_{x}
	\\+&\Delta\rho_{y}\sigma_{y}\tau_{z}+V_{x}\rho_{z}\sigma_{x},
  \end{split}
  \end{equation}
Define new momenta
\begin{equation}
	\tilde{k}_x=\frac{k_x+k_y}{\sqrt{2}}, \quad  \tilde{k}_y=\frac{k_x-k_y}{\sqrt{2}}
\end{equation}
which are parallel to the border of the rhombus-shaped sample in Fig. \ref{fig2}(h).
Through these new momenta, the $H_l$ can be represented as
\begin{equation}
	\begin{split}
		\label{Hlt}
		H_l=&\frac{A}{\sqrt{2}}(\tilde{k}_{x}+\tilde{k}_{y})\sigma_{x}\tau_{z}+
		\frac{A}{\sqrt{2}}(\tilde{k}_{x}-\tilde{k}_{y})\rho_{z}\sigma_{y}\tau_{z}\\+&[m_0-m_1(\tilde{k}_x^2+\tilde{k}_y^2)]\rho_{z}\tau_{x}
		+\Delta_{}\rho_{y}\sigma_{y}\tau_{z}\\+&V_{x}\rho_{z}\sigma_{x},
	\end{split}
\end{equation}
Consider the open boundary condition in $\frac{1}{\sqrt{2}}(1,1)$ direction and periodic boundary condition in
$\frac{1}{\sqrt{2}}(1,-1)$ direction, we also can treat $\tilde{k}_{y}$
term as perturbation. After omitting the high-order perturbation $O(\tilde{k}_{y}^2)$ term,
we have $H_l=H_0+H_p$ with
\begin{equation}
	\begin{split}
	\label{H0}
	H_{0}=&\frac{A}{\sqrt{2}}\tilde{k}_{x}\sigma_{x}\tau_{z}+
	\frac{A}{\sqrt{2}}\tilde{k}_{x}\rho_{z}\sigma_{y}\tau_{z}\\
	+&[m_0-m_1\tilde{k}_x^2]\rho_{z}\tau_{x}+\Delta\rho_{y}\sigma_{y}\tau_{z}
  \end{split}
  \end{equation}
  \begin{equation}
	\begin{split}
	\label{Hp}
	H_{p}=&\frac{A}{\sqrt{2}}\tilde{k}_{y}\sigma_{x}\tau_{z}-
	\frac{A}{\sqrt{2}}\tilde{k}_{y}\rho_{z}\sigma_{y}\tau_{z}+\\
	&V_{x}\rho_{z}\sigma_{x}.
  \end{split}
  \end{equation}
  We treat $H_0$ as unpertubted Hamiltonian and $H_p$ as perturbtation. 
  Notice that $[H_0,\rho_x\sigma_x\tau_z]=[H_0,\rho_x\sigma_y\tau_y]=0$, $\rho_x\sigma_x\tau_z$ has eigenvactors
  \begin{equation}
	\begin{split}
	\label{psi}
	\psi_1=&|\rho_x=1,\sigma_x=1,\tau_z=1\rangle\\
	\psi_2=&|\rho_x=1,\sigma_x=-1,\tau_z=-1\rangle\\
	\psi_3=&|\rho_x=-1,\sigma_x=1,\tau_z=-1\rangle\\
	\psi_4=&|\rho_x=-1,\sigma_x=-1,\tau_z=1\rangle\\
	\psi_5=&|\rho_x=1,\sigma_x=1,\tau_z=-1\rangle\\
	\psi_6=&|\rho_x=1,\sigma_x=-1,\tau_z=1\rangle\\
	\psi_7=&|\rho_x=-1,\sigma_x=1,\tau_z=1\rangle\\
	\psi_8=&|\rho_x=-1,\sigma_x=-1,\tau_z=-1\rangle\\
  \end{split}
  \end{equation}
  $\rho_x\sigma_x\tau_z$ and $\rho_x\sigma_y\tau_y$ have common eigenvactors
  \begin{equation}
	\begin{split}
	\label{psi}
	\chi_1=&\frac{1}{\sqrt{2}}(\psi_1+\psi_2),\quad \chi_2=\frac{1}{\sqrt{2}}(\psi_3+\psi_4)\\
	\chi_3=&\frac{1}{\sqrt{2}}(\psi_1-\psi_2),\quad \chi_4=\frac{1}{\sqrt{2}}(\psi_3-\psi_4)\\
	\chi_5=&\frac{1}{\sqrt{2}}(\psi_5-\psi_6),\quad \chi_6=\frac{1}{\sqrt{2}}(\psi_7-\psi_8)\\
	\chi_7=&\frac{1}{\sqrt{2}}(\psi_5+\psi_6),\quad \chi_8=\frac{1}{\sqrt{2}}(\psi_7+\psi_8)\\
  \end{split}
  \end{equation}
  Using these basis, we can define a unitary transformation
  \begin{equation}
	\begin{split}
	\label{u1}
  U_1=[\chi_1,\chi_2,\chi_3,\chi_4,\chi_5,\chi_6,\chi_7,\chi_8].
  \end{split}
  \end{equation}
  We remark that $\bm{s}$, $\bm{\sigma}$, and $\bm{\tau}$ only represent Pauli matrix in orthognoal space. 
  Consider the rotation of Pauli $\bm{\tau}$ space,
   we can define another unitary transformation
   \begin{equation}
	\begin{split}
	\label{u1}
  U_2=e^{i\pi\tau_z/8}.
  \end{split}
  \end{equation}
  Under these two unitary transformations, we can get new $H_l$, $H_0$, and $H_p$, labeled as 
  $H_l^n$, $H_0^n$, and $H_p^n$, respectively, with 
  \begin{equation}
	\begin{split}
	\label{h0n}
  H_0^n(\tilde{k}_x)=&U_2^{\dagger}U_1^{\dagger}H_0U_1U_2 \\
  =&diag \left[-A\tilde{k}_x\tau_y+(m_0-\Delta-m_1\tilde{k}_x^2)\tau_x, \right.\\
  &A\tilde{k}_x\tau_z+(m_0+\Delta-m_1\tilde{k}_x^2)\tau_x, \\
  &-A\tilde{k}_x\tau_z+(m_0-\Delta-m_1\tilde{k}_x^2)\tau_x, \\
  &\left.A\tilde{k}_x\tau_y+(m_0+\Delta-m_1\tilde{k}_x^2)\tau_x \right],
  \end{split}
  \end{equation}
  \begin{equation}
	\begin{split}
	\label{hpn}
  H_p^n=&U_2^{\dagger}U_1^{\dagger}H_pU_1U_2 \\
  =&\frac{A}{2}\tilde{k}_y(\rho_z\tau_z-\rho_z\tau_y+\sigma_z\tau_z+\sigma_z\tau_y)+V_x\rho_x\tau_x
  \end{split}
  \end{equation}
  and $H_l^n=H_0^n+H_p^n$. Substitute $\tilde{k}_x$ with $-i\partial_{\tilde{x}}$,
   we get the zero modes equation $H_0^n(-i\partial_{\tilde{x}})\phi_{\alpha}=0$. 
  When $|\Delta|>|m_0|$ and $m_1<0$, there are two zero energy solutions on the I edge of Fig. \ref{fig2}(h):
	$\phi_{\alpha}=N_{\tilde{x}}sin(\kappa_1\tilde{x})e^{\kappa_2 \tilde{x}}e^{i\tilde{k}_y\tilde{y}}\xi_{\alpha}$
	\begin{equation}
	  \begin{split}
	  \label{xi}
   \xi_1&=|\rho_z=1,\sigma_z=1,\tau_z=-1 \rangle \\
   \xi_2&=|\rho_z=-1,\sigma_z=1,\tau_y=1 \rangle, 
	\end{split}
	\end{equation}
	$\kappa_1=\frac{\sqrt{4(m_0-\Delta)m_1-A^2}}{2m_1}, \kappa_2=\frac{A}{2|m_1|},$ and 
	$N_{\tilde{x}}=\frac{2\kappa_2 \sqrt{(\kappa_1^2+\kappa_2^2)}}{\kappa_1}$. $\tilde{x}$ and $\tilde{y}$ 
	are the coordinates along $\frac{1}{\sqrt{2}}(1,1)$ and $\frac{1}{\sqrt{2}}(1,-1)$ directions, respectively.
	 Under $\phi_{\alpha}$ $(\alpha=1,2)$, $H_p^n$ can be written as 
	\begin{equation}
	  \begin{split}
	  \label{HI}
   H_I=-A\tilde{k}_y\sigma_z+\frac{V_x}{2}\sigma_x
	\end{split}
	\end{equation}
	In same vein,  we can deduce the effective Hamiltonian on III, II, and IV edges
	\begin{equation}
	  \begin{split}
	  \label{HIIV}
   H_{III}=&A\tilde{k}_y\sigma_z-\frac{V_x}{2}\sigma_x \\
   H_{II}=&A\tilde{k}_x\sigma_z+\frac{V_x}{2}\sigma_x \\
   H_{IV}=&-A\tilde{k}_x\sigma_z-\frac{V_x}{2}\sigma_x
	\end{split}
	\end{equation}
$H_\text{I}, H_\text{II}, H_\text{III},$ and $H_\text{IV}$ are the Jackiw-Rebbi model~\cite{JRPRD1976} describing Dirac equation subjected to a mass kink. Accordingly, we deduce that there are zero mode at the corner of II and III edges as well as
the corner of IV and I edges, as around each of the two corners the Dirac equation on two adjacent edges gain the opposite-sign mass.
\begin{figure}[htbp]
	\centering
	\includegraphics[width=3.2in]{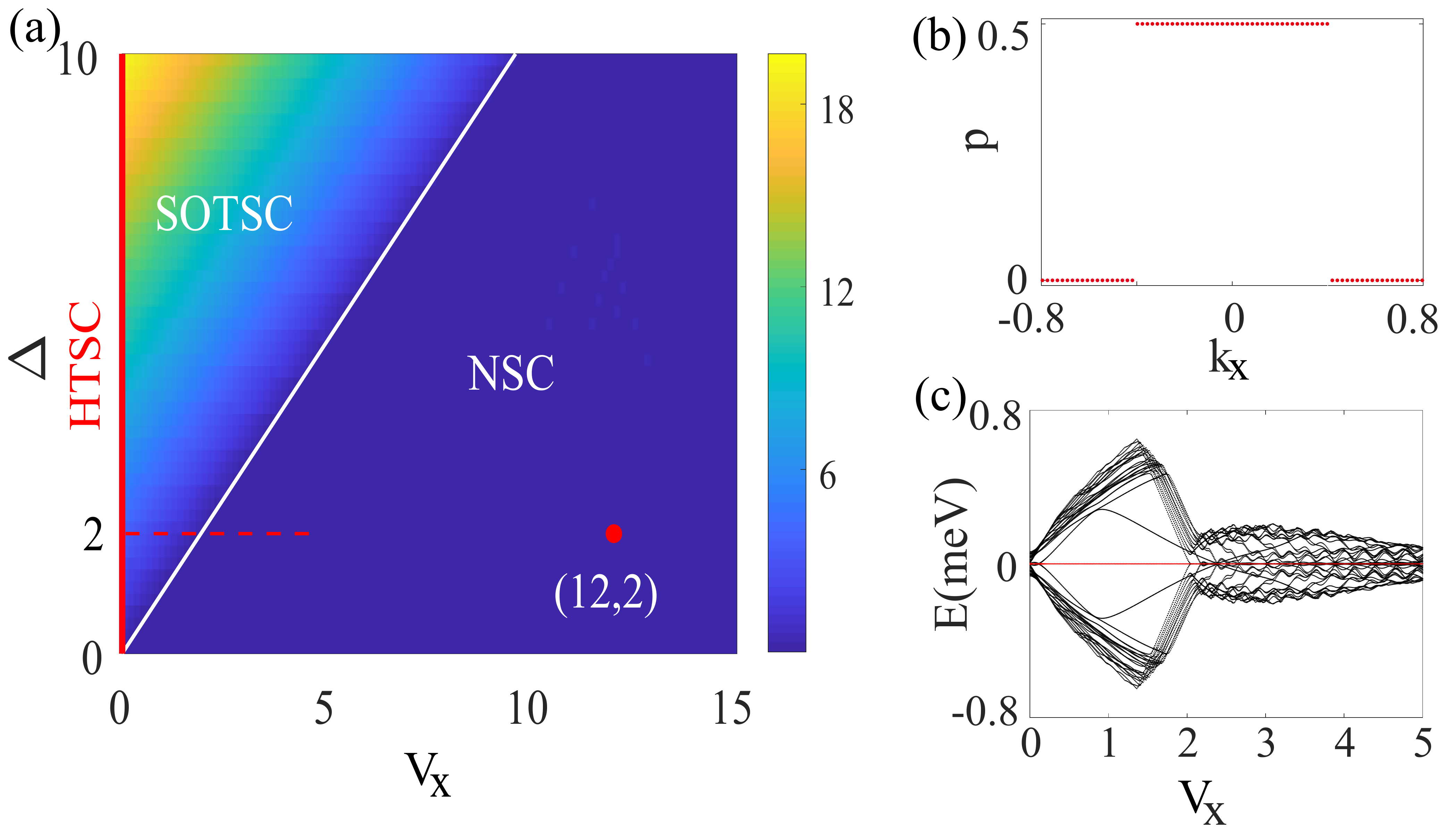}
	\caption{(Color online).
		(a) Phase diagram of topological superconductors. The upper left regime marked by SOTSC represents the second-order topological
		superconductor phase. The lower right regime marked by NSC represents the nodal topological
		superconductor phase. The left red solid line represents the helical topological superconductor (HTSC) phase in the absence of $V_x$. The color bar represents the energy gap. (b) Bulk polarization of the nodal phase as a function of $k_{x}$. 
		In (a) and (b), we fix $\mu=12\,\mathrm{meV}$. In (b), we also fix $\Delta_{t}=2\,\mathrm{meV}$ and $M_{x}=1.5\,\mathrm{meV}$.
		(c) Energy spectrum as a function of $V_x$ for a finite-sized sample with rhombus geometry. The parameters correspond to the red dashed line
		in (a).
		\label{fig3} }
\end{figure}
\begin{figure*}[htbp]
	\centering
	\includegraphics[width=6.0in]{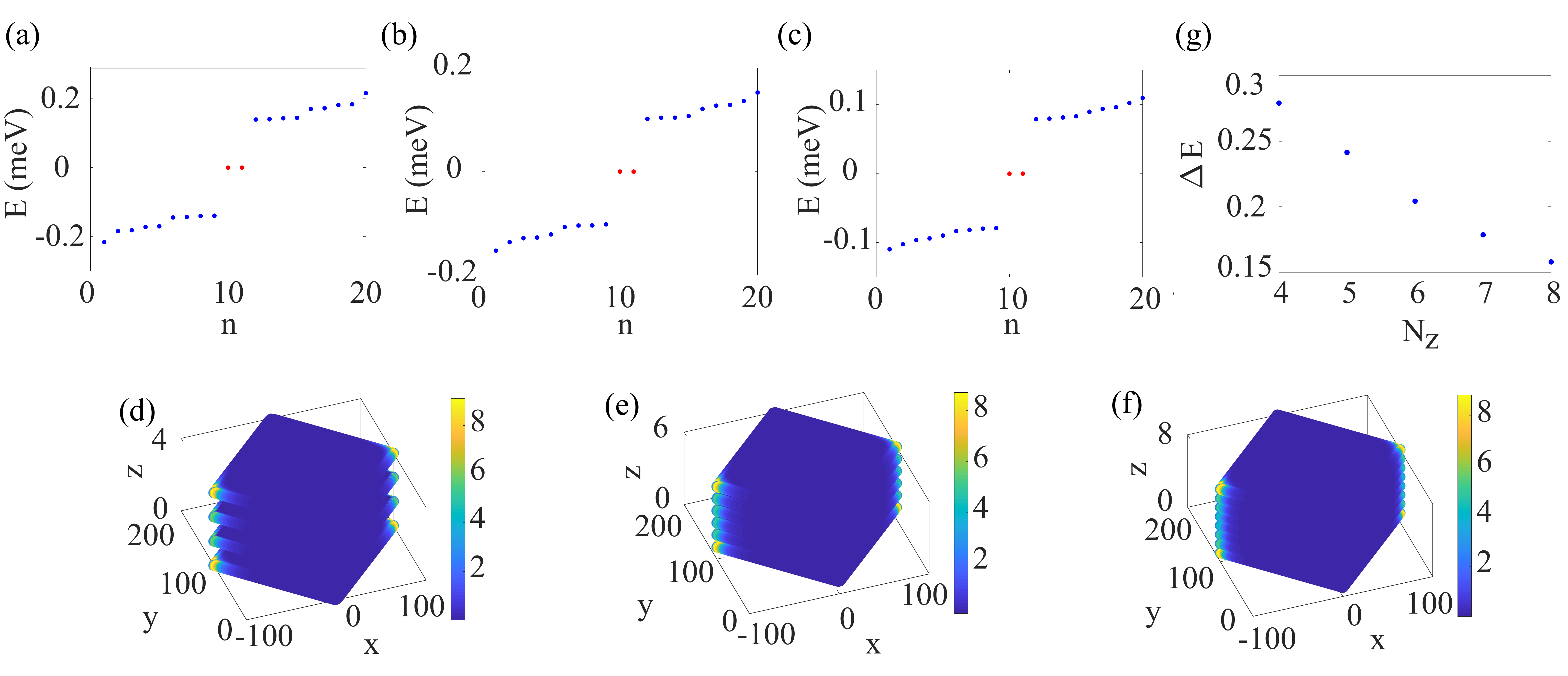}
	\caption{(Color online). $N_{x}\times N_{y}=200\times200$ and rhombus-shaped sample for all plots.
		The energy spectrum for distinct number of layers (a) $N_{z}=4$, (b) $N_{z}=6$, and (c) $N_{z}=8$. (d), (e), and
		(f) are the local density of the two Majorana ingap bound states marked in red dots in (a), (b), and (c), respectively.
		(g) The gap of bulk spectrum marked in blue dots as a function of number of layers $N_{z}$.
		In (a)-(g), {we set $a=5\,\mathrm{nm}$, $A=50\,\mathrm{meV\cdot nm}$
		, $m_{0}=8\,\mathrm{meV}$, $\mu=2\,\mathrm{meV}$,
		$\Delta_{}=2\,\mathrm{meV}$, $V_{x}=1.5\,\mathrm{meV}$, $t_{x}=t_{y}=125\,\mathrm{meV\cdot nm^{2}}$ and $t_{z}=100\,\mathrm{meV\cdot nm^{2}}$}.
		\label{fig4} }
\end{figure*}

Continuing to increase the in-plane Zeeman field $V_{x}$, we find that the gap closes, and a nodal topological superconducting phase with nodes along the $k_{x}$-axis is formed. Considering a ribbon geometry with the open boundary condition along $y$ direction, we plot the energy spectrum in Fig.~\ref{fig2}(c). Clearly, this nodal phase hosts flat-band Majorana edge modes in between the two bulk nodes. The flat-band Majorana edge modes are confirmed by calculating the energy spectrum and the local density of states of the edge modes, displayed in Figs.~\ref{fig2}(f) and \ref{fig2}(i). We can see that the flat-band Majorana edge modes are located on the top and bottom edges of the sample. 

 To capture the topological characteristic of the nodal phase, we further adopt
 the Wilson loop method \cite{berry1984quantal,wilczek1984appearance,alexandradinata2014wilson,benalcazar2017quantized,benalcazar2017electric,franca2018anomalous,luo2019higher}
 to calculate the bulk polarization of the system. Since the bulk nodes are located along the $k_x$-axis, by treating $k_{x}$ as a parameter,
the effective Hamiltonian reduces to a one-dimensional Hamiltonian $H_{k_{x}}(k_{y})$.
The Wilson loop operator \cite{berry1984quantal,alexandradinata2014wilson,wilczek1984appearance,benalcazar2017quantized,benalcazar2017electric,franca2018anomalous,luo2019higher} along the path in the $k_{y}$-direction $W_{y,k_{y}}$
is defined by: $W_{y,k_{y}}=F_{y,k_{y}+(N_{y}-1)\Delta k_{y}}\cdots F_{y,k_{y}+\Delta k_{y}}F_{y,k_{y}}$, where $k_{y}$ is the base point and $N_{y}$ is the number of unit cells in the $y$-direction. Here, $\left[F_{y,k_{y}}\right]^{mn}=\left\langle u_{k_{y}+\Delta k_{y}}^{m}|u_{k_{y}}^{n}\right\rangle $ with the step $\Delta k_{y}=2\pi/N_{y}$, 
and  $|u_{k_{y}}^{n}\rangle$ represents the occupied
Bloch wave functions with $n=1,2,\cdots N_\text{occ}$.
{$N_\text{occ}=N_{b}/2$} is the number of occupied bands, with $N_{b}$ the degrees of freedom for each cell. 
 Fixing $k_{x}$, we
can determine the Wilson loop operator $W_{y,k_{y}}$ on a path along $k_{y}$. The Wannier center $v_{y}^{j}$ can be determined by
 the eigenvalues of the Wilson-loop operator $W_{y,k_{y}}$,
\begin{equation}
\label{Wky}
W_{y,k_{y}}|v_{y,k_{y}}^{j}\rangle=e^{i2\pi v_{y}^{j}}|v_{y,k_{y}}^{j}\rangle,
\end{equation}
where $j\in\left\{1,2,\cdots N_\text{occ}\right\}$ labels eigenstates
$|v_{y,k_{y}}^{j}\rangle$ as well as components $[v_{y,k_{y}}^{j}]^{n}$.
Since $k_{x}$ is fixed, the bulk polarization can be define as
$p=\mathop{\sum_{j}v_{y}^{j}}\,\mathrm{mod}\,1$~\cite{benalcazar2017electric}.
We plot the bulk polarization as a function of $k_{x}$ as shown in Fig. \ref{fig3}(b). It is clear that the polarization is quantized to 1/2
between two nodal points. We remark that the topological characteristic of the nodal phase can be portrayed by the $k_{x}$-dependent
polarization. 

Finally, we plot the topological superconducting phase diagram on the plane of $\Delta$ and $V_{x}$ as shown in Fig.~\ref{fig3}(a).
The phase boundaries is determined by numerically observing the gap closure of the bulk.
The phase boundary between SOTSC and the nodal topological superconductor
(NSC) is {approximately fitted by the line $V_{x}=\Delta_{}$}. By tuning the
in-plane Zeeman field $V_{x}$, SOTSC phase, NSC phase, and helical topological superconductor (HTSC) phase can be achieved.
Figure \ref{fig3}(c) shows the energy spectrum as a function of $V_x$ for the finite-sized sample with rhombus-like geometry in Fig.~\ref{fig2}(h).

\section{The effect of thickness on Majorana corner modes} \label{seciv}
In this section, we study how the thickness of the topological insulator thin film affect the Majorana corner modes. 
In the three dimensional limit, the bulk Hamiltonian of the intermediate topological insulator can be expressed as:
\begin{equation}
  \begin{split}\label{Htd1}
      H_\text{3DTI}({\bf k})=
      &\frac{A}{a}\sin(k_{x}a)\sigma_{x}\tau_{x}+\frac{A}{a}\sin(k_{y}a)\sigma_{y}\tau_{x}-\mu\\
      +&\frac{A}{a}\sin (k_{z}a)\sigma_{z}\tau_{x}+ M({\bf k})\tau_{z}+
      V_{x}\sigma_{x},
  \end{split}
  \end{equation}
where $M({\bf k})=m_{0}-2\frac{t_{x}}{a^2}(1-\cos (k_{x}a))-2\frac{t_{y}}{a^2}(1-\cos (k_{y}a))-2\frac{t_{z}}{a^2}(1-\cos (k_{z}a))$ with ${\bf k}=(k_x,k_y,k_z)$, and $V_{x}$ is the Zeeman field lies along the $x$-direction.
For simplicity, we consider the proximity-induced superconducting gap only exits on outermost layers of the top and bottom surfaces of the three dimensional topological insulator, and the top and bottom superconductors remain a $\pi$ phase shift.
 In reality, the proximity-induced superconducting potential decreases exponentially and can extend to several layers~\cite{wang2012coexistence,xu2014artificial}. However, this will not change the physics discussed in this section. 
Considering the confinement of the topological insulator thin film along the $z$-direction,
the total Hamiltonian in the Nambu space can be written as,
\begin{equation}
\begin{split}
\label{Htd2}
    H_\text{3D}({\bf k_{\parallel}})=&\sum_{z=1}^{N_z-1}b_{{\bf k}_{\parallel},z}^{\dagger}
    (-\frac{iA}{2}\sigma_{z}\tau_{x}+t_z\rho_{z}\tau_{z})b_{{\bf k}_{\parallel},z+1}+\text{h.c.}
    \\+&\sum_{z=1}^{N_z}b_{{\bf k}_{\parallel},z}^{\dagger}[A\sin (k_{x}a)\sigma_{x}\tau_{x}+A\sin (k_{y}a)\rho_{z}\sigma_{y}\tau_{x}
    \\+&V_{x}\rho_{z}\sigma_{x}-\mu\rho_{z}+\tilde{M}({\bf k_{\parallel}})\rho_{z}\tau_{z}]b_{{\bf k}_{\parallel},z}
    \\+&(b_{{\bf k}_{\parallel},1}^{\dagger}\Delta_{}\rho_{y}\sigma_{y}b_{{\bf k}_{\parallel},1}
    -b_{{\bf k}_{\parallel},N_z}^{\dagger}\Delta_{}\rho_{y}\sigma_{y}b_{{\bf k}_{\parallel},N_z}),
\end{split}
\end{equation}
where $b_{\mathbf{k}_{\parallel},z}=[\psi_{l,\alpha}(\mathbf{k}_{\parallel},z),\psi_{l,\alpha}^{\dagger}(-\mathbf{k}_{\parallel},z)]^{T}$ with the orbital index $l=P1$ $(P2)$ and the spin index $\alpha=\uparrow$ ($\downarrow$). 
$\tilde{M}({\bf k}_{\parallel})=m_{0}-2t_{x}(1-\cos (k_{x}a))-2t_{y}(1-\cos (k_{y}a))-2t_z/a^2$
and ${\bf k}_{\parallel}=(k_x,k_y)$.
Generally, the low energy physics of the topological insulator thin film $H_\text{3D}({\bf k_{\parallel}})$
can be described by the 2D Hamiltonian $H({\bf k})$ in Eq.~(\ref{Htd}) \cite{Liu2010Oscillatory}.
Similar to $H({\bf k})$, the Hamiltonian $H_\text{3D}({\bf k_{\parallel}})$ preserves the mirror-like symmetry:
$C_{2y}H_\text{3D}(k_x,k_y)C_{2y}^{-1}=H_\text{3D}(-k_x,k_y)$ with $C_{2y}=i\sigma_x{\mathcal U}$.
Here $N_z\times N_z$ antidiagonal matrix ${\mathcal U}_{ij}=\delta_{i+j,N_z+1}$  is defined in real
space with Kronecker delta $\delta_{i,j}$.
Thereore, it is natural to expect the Majorana corner modes will be sustainable with the varying thickness.
For $\Delta \neq0$, $V_{x}=0$, the system is a helical topological superconductor, if the Fermi energy is outside the surface gap \cite{liu2011helical}. By turning on the Zeeman field $V_{x}>0$, the gapless Majorana corner modes are observed, as shown in Fig.~\ref{fig4}. 
Figures \ref{fig4}(a), \ref{fig4}(b), and \ref{fig4}(c) show the energy spectrum
(only the 20 eigenenergies with the smallest absolute value are shown)
 of an $N_{x}\times N_{y}=200\times200$
rhombus-shaped sample. Again, two zero-energy ingap Majorana bound states appear (marked by the red dots in Fig. \ref{fig4}).
Figures \ref{fig4}(d), \ref{fig4}(e), and \ref{fig4}(f) are the local density of the two zero-energy Majorana bound states in
Figs.~\ref{fig4}(a), \ref{fig4}(b), and \ref{fig4}(c), respectively. We can see that the two Majorana bound states are also localized at two opposite corners in the $xy$-plane, but extended in hinge of the side surface along the $z$-direction. In order to explore the
relationship between the gap of side-surface spectrum [marked in blue dots in Figs.~\ref{fig4}(a)-\ref{fig4}(c)]
and the number of layers $N_{z}$, we plot the band gap for different number of layers $N_{z}$ in Fig.~\ref{fig4}(g). It implies
that the band gap decreases rapidly with the increase of  $N_{z}$ due to the decreasesing finite-size confinement along the $z$-direction.

\section{CONCLUSIONS} 
\label{secv}

In conclusion, we have illustrated that an SOTSC with two Majorana corner modes is realized in topological insulator thin film based superconducting junctions with a $\pi$ phase shift when an in-plane Zeeman field is applied. We employ the mirror winding number to characterize the second-order topology of Majorana corner modes. We also analytically deduce an edge theory for the Majorana corner modes by using the perturbation theory. By tuning the Zeeman field, we also observe a nodal superconducting phase hosting flat-band Majorana edge modes, whose bulk topology can be captured by a $k$-dependent polarization. At last, we demonstrate that how the thickness of topological insulator thin films affects the Majorana corner modes and their spatial distribution.

\section*{Acknowledgments} 
\label{secvi}
The authors acknowledge the support by the NSFC (under
Grants No. 12074108, No. 11974256, and No.12147102), and the
Priority Academic Program Development (PAPD) of Jiangsu
Higher Education Institution.

\bibliography{MajorHOTSC}

\end{document}